\documentclass[reprint,
 amsmath,amssymb,
 aps,
floatfix,
showkeys
]{revtex4-1}

\usepackage[USenglish]{babel}
\usepackage{slashed}
\usepackage{physics}
\usepackage{graphicx}
\usepackage{dcolumn}
\usepackage{bm}
\usepackage[usenames]{color}
\usepackage{xcolor}
\usepackage{float}

\newcommand{\be}{\begin{equation}}
\newcommand{\ee}{\end{equation}}
\newcommand{\bea}{\begin{eqnarray}}
\newcommand{\eea}{\end{eqnarray}}
\newcommand{\beas}{\begin{eqnarray*}}
\newcommand{\eeas}{\end{eqnarray*}}
\newcommand{\nn}{\nonumber\\}

\begin{document}

\title{Effective potential and mass behavior of a self-interacting scalar field theory due to thermal and external electric and magnetic fields effects}


\author{M. Loewe$^{1,2,3,4}$, D. Valenzuela$^{1}$ and R. Zamora$^{5,6}$ }
\affiliation{%
$^1$Instituto de F\'isica, Pontificia Universidad Cat\'olica de Chile, Casilla 306, Santiago 22,Chile.\\
$^2$Centre for Theoretical and Mathematical Physics, and Department of Physics, University of Cape Town, Rondebosch 7700, South Africa.\\
$^3$Facultad de Ingeniería, Arquitectura y Diseño, Universidad San Sebastián, Santiago, Chile.\\
$^4$Centro Cient\'ifico-Tecnol\'ogico de Valpara\'iso CCTVAL, Universidad T\'ecnica Federico Santa Mar\'ia, Casilla 110-V, Valpara\'iso, Chile.\\
$^5$Instituto de Ciencias B\'asicas, Universidad Diego Portales, Casilla 298-V, Santiago, Chile.\\
$^6$Centro de Investigaci\'on y Desarrollo en Ciencias Aeroespaciales (CIDCA),  Academia Politécnica Aeronáutica, Fuerza A\'erea de Chile, Casilla 8020744, Santiago, Chile.}%

\begin{abstract}

    In this article we address the subject of finding the behavior of a charged scalar field $\phi$ under the influence of external constant magnetic and electric fields, perpendicular to each other, including also thermal effects. For this purpose we derive an expression for the corresponding bosonic propagator. As an application, we explore, in the weak field sector, the mass correction for the self interacting $\lambda \phi ^4$ theory. Our results show that the mass diminishes when the magnetic field appears, for small values of temperature, staring to increase then when the strength of the field rises. In the case  when we have only an electric field, the mass always grow with the field intensity. We also analyze the phase diagram associated to spontaneous symmetry breaking of the theory finding inverse magnetic catalysis (IMC) or inverse electric catalysis (IEC) for the cases where only a magnetic field or only an electric field are present, respectively. In both cases, taken separately,  we have a scenario where the critical temperature associate to symmetry restoration diminishes as function of the corresponding field strengths. A similar situation happens when both type of fields are simultaneously present. We have dubbed this case as inverse magnetic -electric catalysis (IMEC). In this situation, both fields cooperate for the occurrence of IMEC.
    
\end{abstract}



\maketitle

\section{Introduction}\label{sec1}

In this article we derive a propagator for a charged scalar field in the presence of constant external magnetic and electric fields. This propagator is used then to explore the mass behavior of the $\phi $ field, in the frame of a self-interacting $\lambda \phi ^4$ theory, as function of temperature and the external magnetic and electric field strengths, taking both fields simultaneously.  As a second application, we analyze the phase diagram of this theory starting from the broken phase. For this purpose we find the effective potential, including the resummation of ring diagrams, finding the dependence of the critical temperature on
the external electromagnetic field strengths. 

 In the existing literature, different aspects concerning the behavior of hadronic parameters and the properties of the QCD phase diagram in the presence of thermal and external magnetic effects have been considered during the last years \cite{iranianos,zamora1,zamora2,zamora3,simonov03,aguirre02,tetsuya,dudal04,kevin,gubler,noronha01,morita,Ayala1,morita02,sarkar03,band,nosso1,nosso03,Ayala2,zamora4,peng,cohen,tavares,inverse3,inverse4,inverse5,inverse6,inverse7}. In \cite{reviews} we might find general review articles concerning the role of magnetic effects in strongly interacting matter. In particular, it is interesting to mention that it is possible to find inverse magnetic catalysis in the frame of effective models, in agreement with the well known results from the lattice community \cite{lattice}, i.e. the fact that the critical temperature diminishes with the magnetic field strength. We have analyzed this point by a careful discussion of symmetry restoration both in a self-interacting scalar field model as well as in the linear sigma field coupled to quarks, dealing in the last case with chiral symmetry restoration. It turns out that this can only be  achieved by going beyond the mean field approximation, through a ring diagram resummation. \cite{nosotros1,nosotros2}. A recent review about these facts, including a discussion of thermo-magnetic effects in the Nambu-Jona-Lasinio model, is presented in \cite{cristian}. A discussion of the effective potential, in the case where we have a pure external electric field, shows the occurrence of electric anticatalysis for small values of the electric field, whereas catalysis appears for high values of the electric field strength \cite{Farias,ultimonosotros}. We also want to mention, in the context of the influence of external fields, a complete different aspect, namely the evolution of the residues of renormalons in this theory under the influence of an electric field which has also recently  been addressed by some of us. \cite{renormalon2}.

In this article we would like to consider the situation or scenario where the effects of both kind of external fields, electric and magnetic, are taken into account. The physical scenario where such a case appears corresponds to peripheral collisions of asymmetric heavy ions. For example, in Au-Cu collisions an electric dipole type field is generated due to the imbalance of the number of charged particles (protons) in both nuclei. This field will basically be perpendicular to the magnetic field generated also under such conditions. In \cite{paralelos} we might find an analysis when both types of fields are present, but in a parallel configuration, which corresponds actually to the chiral magnetic effect. For the purpose of analyzing collisions between asymmetric nuclei, however, the configuration where both fields are perpendicular to each other is the relevant one. 


\noindent
This article is organized as follows. In section \ref{sec2} we derive the bosonic propagator in the presence of external constant magnetic and electric fields, in a perpendicular configuration relative to each other. We also present the propagator's  expansion for the case of weak fields. In section \ref{sec3} we make use of the propagator to find the electromagnetic mass correction of the charged scalar field. In section  \ref{sec4} we proceed with the discussion of the effective potential at the one loop level, including also ring corrections, finding the behavior of the  critical temperature where the symmetry is restored as function of the magnetic and electric field strengths. Finally, in section \ref{sec5} we present our conclusions.
\section{The boson propagator in the presence of an electric and magnetic field}\label{sec2}

\noindent We begin with the proper time representation for a charged boson propagator in the presence of an external electromagnetic field encoded in the $F^{\mu \nu}$ tensor \cite{Dittrich,ahmad} which is given by

\begin{equation}
D(p)=\displaystyle{\int}_0^{\infty}dt\,e^{-m^2t} \frac{e^{tp_{\mu}\left(\frac{\tan(Z)}{Z}\right)^{\mu\nu}p_{\nu}}}{\sqrt{\det(\cos(Z))}}, \label{pro}
\end{equation}
where $Z^{\mu\nu}\equiv qF^{\mu\nu}t$, with $q$ the electric charge and $p$ the euclidean four-momentum defined as $p=(p_1,p_2,p_3,p_4)$. Our analysis is presented in the Euclidean formulation.\\
As we said in the introduction, we are interested in a configuration where $\Vec{B}\perp\Vec{E}$. Without loss of generality, we assume that $\Vec{B}=B_0\hat{z}$ and $\Vec{E}=E_0\hat{x}$ that is\\
\begin{equation}
    F^{\mu\nu}=\left[\begin{array}{cccc}
         0&B_0&0&i E_0  \\
         -B_0&0&0&0\\
         0&0&0&0\\
         -i E_0&0&0&0
    \end{array}\right].
\end{equation}
Thus, we can decompose $F^{\mu\nu}$ as\\
\begin{equation}
    F^{\mu\nu} \equiv iE_0E^{\mu\nu}+B_0B^{\mu\nu},
\end{equation}where\\
\begin{eqnarray}
    E^{\mu\nu} &=&\begin{bmatrix}
    0&0&0&1\\
    0&0&0&0\\
    0&0&0&0\\
    -1&0&0&0
    \end{bmatrix}  \mbox{and}\\
    B^{\mu\nu} &=&\begin{bmatrix}
        0&1&0&0\\
        -1&0&0&0\\
        0&0&0&0\\
        0&0&0&0
    \end{bmatrix}.
\end{eqnarray}
Let us compute the square of $F^{\mu\nu}$
\begin{equation}
    (F^2)^{\mu\nu} = E_0^2(E^2)^{\mu\nu}+B_0^2(B^2)^{\mu\nu}+iE_0B_0 S^{\mu\nu},
\end{equation}with\\
\begin{eqnarray}
(E^2)^{\mu\nu}&=& \begin{bmatrix}
     1&0&0&0\\
     0&0&0&0\\
     0&0&0&0\\
     0&0&0&1
 \end{bmatrix},\\
 (B^2)^{\mu\nu}&=& \begin{bmatrix}
     -1&0&0&0\\
     0&-1&0&0\\
     0&0&0&0\\
     0&0&0&0
 \end{bmatrix} \mbox{and}\\
 S^{\mu\nu}&=& \begin{bmatrix}
     0&0&0&0\\
     0&0&0&-1\\
     0&0&0&0\\
     0&-1&0&0
 \end{bmatrix},
\end{eqnarray}
We find the following relations for the powers of $F^{\mu\nu}$\\
\begin{eqnarray}
    (F^{2n})^{\mu\nu} &=&\left[-\left(-E_0^2+B_0^2\right)\right]^{n-1}(F^2)^{\mu\nu}\\
    (F^{2n+1})^{\mu\nu} &=&\left[-\left(-E_0^2+B_0^2\right)\right]^n\,F^{\mu\nu}\,\,\forall \, n \in \mathbb{N}^0.
    \end{eqnarray}
    Summarizing, we have\\
    \begin{eqnarray}
 (F^0)^{\mu\nu} &\equiv &\delta^{\mu\nu},\\
  (F^{2n})^{\mu\nu} &=&\left(-\lambda^2\right)^{n-1}(F^2)^{\mu\nu},\\
  (F^{2n+1})^{\mu\nu} &=&\left(-\lambda^2\right)^n\,F^{\mu\nu}\,\,\forall \, n \in \mathbb{N}^0,
    \end{eqnarray}
       where we have defined\\
   \begin{equation}
       \lambda^2\equiv -E_0^2+B_0^2.
          \end{equation}
    With these relations, we obtain\\
    \begin{eqnarray}
        \left[\cos(Z)\right]^{\mu\nu}&=&\delta^{\mu\nu}+2\frac{\sin^2(\frac{1}{2}q\lambda it)}{\lambda^2}\,(F^2)^{\mu\nu}\nonumber \\
        &\Rightarrow & \det[\cos(Z)]=\cos^2(q\lambda it), \label{det}
    \end{eqnarray} 
    and
    \begin{equation}
        \left[\frac{\tan(Z)}{Z}\right]^{\mu\nu}=\delta^{\mu\nu}-\left[\frac{ \tan(q\lambda it)}{q\lambda it}-1\right]\frac{(F^2)^{\mu\nu}}{\lambda^2}. \label{tan}
    \end{equation}
Finally, inserting  Eq. (\ref{det}) and Eq. (\ref{tan}) in Eq. (\ref{pro}), we get\\
\begin{eqnarray}
    D(p)&=&\displaystyle{\int}_0^{\infty}dt e^{-t\left(p_3^2+\frac{1}{\lambda^2}(B_0p_4-i E_0p_2)^2+m^2\right)}\nonumber \\
    &\times&\frac{e^{\frac{-\tan(q\lambda it)}{q\lambda i}\left(p_1^2+\frac{1}{\lambda^2}\left(B_0p_2+i E_0 p_4\right)^2\right)}}{\cos(q\lambda it)} \nonumber \\
    &=&\displaystyle{\int}_0^{\infty}dt e^{-t\left(p_3^2+\frac{1}{\lambda^2}(B_0p_4-i E_0p_2)^2+m^2\right)}\nonumber \\
    &\times&\frac{e^{\frac{-\tanh(q\lambda t)}{q\lambda }\left(p_1^2+\frac{1}{\lambda^2}\left(B_0p_2+i E_0 p_4\right)^2\right)}}{\cosh(q\lambda t)}\nonumber \\
    &=&\displaystyle{\int}_0^{\infty}dt e^{-t\left(p_3^2+\frac{1}{-E_0^2+B_0^2}(B_0p_4-i E_0p_2)^2+m^2\right)}\nonumber \\
    &\times&\frac{e^{\frac{-\tanh(q \sqrt{-E_0^2+B_0^2} t)}{q\sqrt{-E_0^2+B_0^2}}\left(p_1^2+\frac{1}{-E_0^2+B_0^2}\left(B_0p_2+i E_0 p_4\right)^2\right)}}{\cosh(q\lambda t)}. \nonumber \\ \label{prop}
\end{eqnarray}

 \subsection{Weak field expansion of the propagator}
 In order to find a weak field expansion for the propagator, we follow the ideas presented in  \cite{ayalaeuro},  and proceed with an expansion for small $E_0$ and $B_0$ values in Eq. \eqref{prop} using the expressions
 \begin{align}
     \frac{\tanh(x)}{x} &\approx 1-\frac{1}{3}x^2+\frac{2}{15}x^4\\
     \sech(x)&\approx 1-\frac{1}{2}x^2\\
     e^x&\approx 1+x .
 \end{align}
After integrating the above expression in the proper time $t$, we obtain the expansion of the propagator given in Eq. \eqref{prop} up to the order $\mathcal{O}(E^2)$ and $\mathcal{O}(B^2)$. We get  \\
 \begin{eqnarray}
 D(p)&\approx&\frac{1}{p^2+m^2}+\frac{4ip_4p_2 (qB)(qE)}{(p^2+m^2)^4}\nonumber \\
 &+&\frac{(qE)^2(m^2+p^2-2(p_4^2+p_1^2))}{(p^2+m^2)^4}\nonumber \\
 &-&\frac{(qB)^2(m^2+p^2-2p_{\perp}^2)}{(p^2+m^2)^4},  
 \end{eqnarray}
 where $p_{\perp}=(0, p_{1},p_{2},0)$. It is important to avoid $p_4^2$ terms in the numerator of the above expression. When going into a finite temperature scenario, those terms are cumbersome to be handled. We can write the previous expression as 
 \begin{eqnarray}
 D(p)&&\approx  \frac{1}{p^2+m^2}+\frac{4ip_2 p_4 (qB) (qE)}{(p^2+m^2)^4} \nonumber \\
&-& \frac{(qE)^2+(qB)^2}{(p^2+m^2)^3} +  2\frac{(qB)^2p_{\perp}^2+(qE)^2(p_2^2+p_3^2)}{(p^2+m^2)^4} \nonumber \\
&+&\frac{2 (qE)^2 m ^2}{(p^2+m^2)^4}.   \label{debil}
\end{eqnarray}

 \section{Mass correction}\label{sec3}
\begin{figure}[h]
\includegraphics[width=6cm]{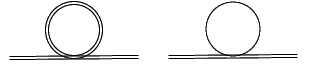} 
\caption{Feynman diagrams that contribute to the self-energy of the $\sigma$ field. The solid line corresponds to the $\chi$ field and the double line to the   $\sigma$ field.}
\label{autoenergiachi}
\end{figure}

\begin{figure}[h]
\includegraphics[width=8.8cm]{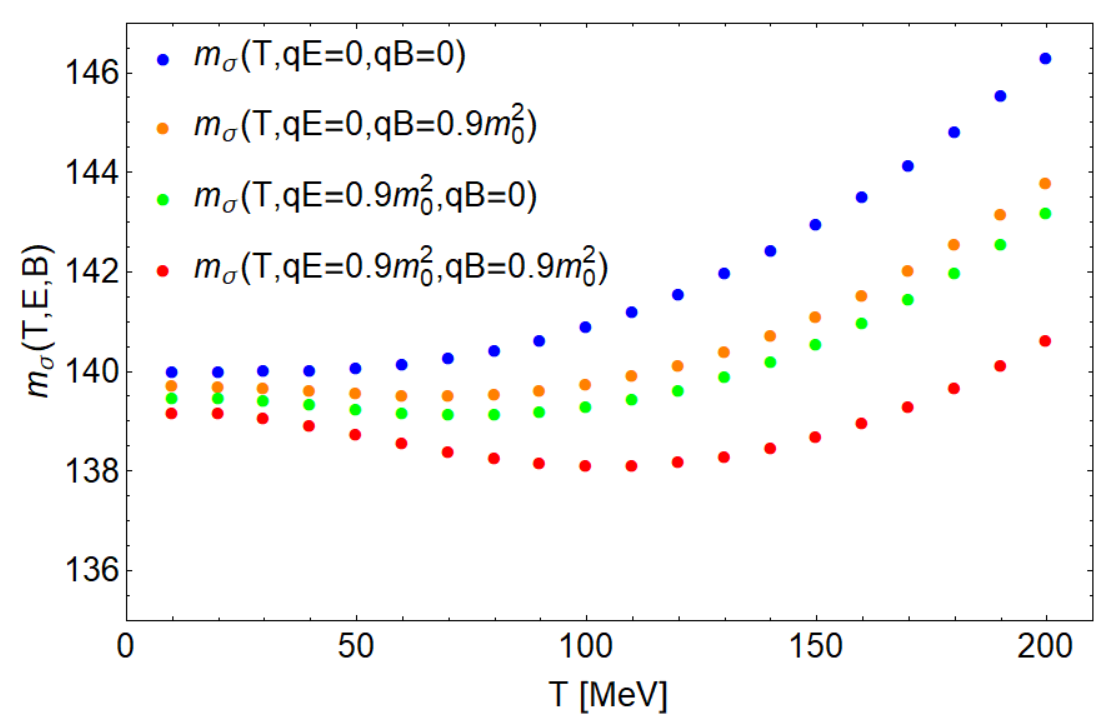} 
\caption{Color on-line. Sigma mass as function of temperature for different strengths of the electric and/or magnetic fields.}
\label{masavst}
\end{figure}

There are several quantities that could be analyzed using the propagator derived in the previous section. Perhaps, the most simple situation is the analysis of the electromagnetic mass correction, including also temperature effects valid for the whole range of temperature.  Next, we  will consider the effective potential, including ring contributions. We remember that we are dealing here with the $\lambda \phi ^4$ theory,
\bea
   {\mathcal{L}}=(D_{\mu}\phi)^{\dag}D^{\mu}\phi
   +\mu^{2}\phi^{\dag}\phi-\frac{\lambda}{4}   
   (\phi^{\dag}\phi)^{2},
\label{lagrangian}
\eea
where $\phi$ is a charged scalar field and
 \bea
   D_{\mu}=\partial_{\mu}+iqA_{\mu}.
\label{dcovariant}
\eea
The squared mass parameter $\mu^2$ and the self-coupling $\lambda$ are taken to be positive. Notice that we are in the broken phase.
We can write the complex field $\phi$ in terms of real components $\sigma$ and $\chi$,
\bea
   \phi(x)&=&\frac{1}{\sqrt{2}}[\sigma(x)+i \chi(x)],  \nonumber \\
   \phi^{\dag}(x)&=&\frac{1}{\sqrt{2}}[\sigma(x)-i\chi(x)].
\label{complexfield}
\eea
Following the usual procedure that allows the occurrence of spontaneous symmetry breaking, the $\sigma$ field develops a vacuum expectation value $v$
\bea
   \sigma \rightarrow \sigma + v,
\label{shift}
\eea
which can later be taken as the order parameter of the theory. After this shift, the Lagrangian can be rewritten as
\bea
   {\mathcal{L}} &=& -\frac{1}{2}[\sigma(\partial_{\mu}+iqA_{\mu})^{2}\sigma]-\frac{1}
   {2}\left(\frac{3\lambda v^{2}}{4}-\mu^{2} \right)\sigma^{2}\nn
   &-&\frac{1}{2}[\chi(\partial_{\mu}+iqA_{\mu})^{2}\chi]-\frac{1}{2}\left(\frac{\lambda v^{2}}{4}-   
   \mu^{2} \right)\chi^{2}+\frac{\mu^{2}}{2}v^{2}\nn
  &-&\frac{\lambda}{16}v^{4}
  +{\mathcal{L}}_{I},
  \label{lagranreal}
\eea
where ${\mathcal{L}}_{I}$ is given by
\bea
  {\mathcal{L}}_{I}&=&-\frac{\lambda}{16}\left(\sigma^4+\chi^4+2\sigma^2\chi^2\right),
  \label{lagranint}
\eea
From Eq.~(\ref{lagranreal}) we see that the $\sigma$ and $\chi$ masses are given by
\bea
  m^{2}_{\sigma}&=&\frac{3}{4}\lambda v^{2}-\mu^{2},\nn
  m^{2}_{\chi}&=&\frac{1}{4}\lambda v^{2}-\mu^{2}.
\label{masses}
\eea
For the discussion of the propagator, we will consider only the $\sigma$ boson mass correction which is given by
\begin{equation}
    m_{\sigma}(T,B,E)=m_0+\Pi_{\sigma}(T,B,E) \label{masacorregida}, 
\end{equation} 
where $m_0$ is the bare mass of the  $\sigma$ field, being  $m_{\sigma}(T,B,E)$  the mass at the one loop level, corrected by temperature and the external electromagnetic fields.
The Feynman diagrams that contribute to the self-energy of the  $\sigma$ boson are shown in figure Fig. (\ref{autoenergiachi}).
Therefore, the self-energy $\Pi_{\sigma}(T,B,E)$ corresponds to 
\begin{eqnarray}
\Pi_{\sigma}(T,B,E)=\frac{\lambda}{4}(12 \Pi(m_\sigma)+2\Pi(m_\chi)), \label{autoen}
\end{eqnarray}
where
\begin{equation}
  \Pi(m_i)\equiv\Pi=T\sum_n \int \frac{d^3p}{(2\pi)^3} D(\omega_n,p,m_i), \label{auto}
\end{equation}
being $m_i$ the $\sigma$ or the $\chi$ mass. Finite temperature effects are handled in the imaginary time formalism in the usual way, i.e.
\begin{equation}
p_4 \rightarrow \omega_{n} = 2\pi n T,\; n\in Z.
\end{equation} 
 where the integral in $p_4$ converts into a sum over  Matsubara frequencies according to,
\begin{eqnarray}
\int \frac{d^4p}{(2\pi)^4}f(p) \rightarrow T\sum_{n\in Z} \int \frac{d^3p}{(2\pi)^3} f(\omega_n,p),
\end{eqnarray}
where we have introduced the notation $p=(p_1,p_2,p_3)$. To calculate Eq. (\ref{auto}) in an analytic way, we proceed through the weak field expansion of the propagator, up to a quadratic order of Eq. (\ref{debil}), getting
\begin{eqnarray}
\Pi&=& T\sum_n \int \frac{d^3p}{(2\pi)^3} \Biggl[ \frac{1}{\omega_n^2+p^2+m^2}+\frac{4ip_2 \omega_n (qB) (qE)}{(\omega_n^2+p^2+m^2)^4} \nonumber \\
&-& \frac{(qE)^2+(qB)^2}{(\omega_n^2+p^2+m^2)^3} +  2\frac{(qB^2)p_{\perp}^2+(qE)^2(p_2^2+p_3^2)}{(\omega_n^2+p^2+m^2)^4} \nonumber \\
&+&\frac{2 (qE)^2 m ^2}{(\omega_n^2+p^2+m^2)^4}  \Biggr] \nonumber \\
&\equiv& \Pi_{\text{I}}+\Pi_{\text{II}}+\Pi_{\text{III}}+\Pi_{\text{IV}}+\Pi_{\text{V}},
\end{eqnarray}
where
\begin{eqnarray}
    \Pi_{\text{I}}&=&T\sum_n \int \frac{d^3p}{(2\pi)^3} \frac{1}{\omega_n^2+p^2+m^2} \nonumber \\
    \Pi_{\text{II}}&=&T\sum_n \int \frac{d^3p}{(2\pi)^3}\frac{4ip_2 \omega_n (qB) (qE)}{(\omega_n^2+p^2+m^2)^4} \nonumber \\
    \Pi_{\text{III}}&=& -T\sum_n \int \frac{d^3p}{(2\pi)^3}\frac{(qE)^2+(qB)^2}{(\omega_n^2+p^2+m^2)^3} \nonumber \\
    \Pi_{\text{IV}}&=&T\sum_n \int \frac{d^3p}{(2\pi)^3}2\frac{(qB)^2p_{\perp}^2+(qE)^2(p_2^2+p_3^2)}{(\omega_n^2+p^2+m^2)^4} \nonumber \\
    \Pi_{\text{V}}&=& T\sum_n \int \frac{d^3p}{(2\pi)^3}\frac{2 (qE)^2 m ^2}{(\omega_n^2+p^2+m^2)^4}.
\end{eqnarray}
The first term in the above expression corresponds just to the vacuum plus a thermal contribution. The second term vanishes, due to symmetry reasons, when doing the integral in  $p_2$. Then we have a sequence of a pure magnetic term followed by a pure electric contribution, a thermomagnetic contribution and, finally, a thermoelectric term. Let us first calculate  $\Pi_{\text{I}}$,
\begin{eqnarray}
&&\Pi_{\text{I}}=T\sum_n \int \frac{d^3p}{(2\pi)^3} \frac{1}{\omega_n^2+p^2+m^2} \nonumber \\
&=&\frac{1}{2\pi^2}\int_0^{\infty} \frac{p^2}{2 \sqrt{p^2+m^2}}(1+2n_B(\sqrt{p^2+m^2})). \label{auto1}
\end{eqnarray}
We do first the sum over Matsubara frequencies according to 
\cite{LeBellac,Kapusta} 
\begin{eqnarray}
T\sum_n \frac{1}{(\omega_n^2+p^2+m^2)}&=& \frac{1}{2 \sqrt{p^2+m^2}} \nonumber \\
&\times&(1+2n_B(\sqrt{p^2+m^2})), \label{suma}
\end{eqnarray}
where $n_B$ is the  Bose-Einstein distribution
\begin{equation}
 n_B(x)=\frac{1}{e^{x/T}-1}.   
\end{equation}
The first term in Eq. (\ref{auto1}) is handled through the standard renormalization procedure in the $\overline{\text{MS}}$ scheme, and we obtain
\begin{eqnarray}
\frac{1}{2\pi^2}\int_0^{\infty} \frac{p^2}{2 \sqrt{p^2+m^2}}=-\frac{m_i^2}{16\pi^2} \biggl(
\ln\left(\frac{\widetilde{\mu}^2}{m_i^2}\right)-\frac{1}{2} \biggr),\nonumber \\
\end{eqnarray}
where $\widetilde{\mu}$ is the ultraviolet renormalization scale. The second term in Eq. (\ref{auto1}) is 
\begin{eqnarray}
  &&   \frac{1}{2\pi^2} \int_0^{\infty} dp \frac{p^2}{\sqrt{p^2+m_0^2}}n_B(\sqrt{p^2+m^2}) \nonumber \\
    &=&\frac{1}{2\pi^2}\sum_{n=0}^{\infty}\int_0^{\infty} \frac{p^2e^{-(n+1)\sqrt{p^2+m^2}/T}}{\sqrt{p^2+m^2}} \nonumber \\
    &=& \frac{mT}{2 \pi^2} \sum_{n=1}^{\infty}\frac{K_1(nm/T)}{n},
\end{eqnarray}
where the modified Bessel function $K_{\alpha}(x)$ has the form \cite{abramowitz}
\begin{equation}
 K_{\alpha}(x)=\int_0^{\infty}du e^{-x \cosh(u)}\cosh(\alpha u).\label{bessel}   \end{equation}
It is easy to see that $ \Pi_{\text{II}}$ vanishes. Let us proceed now to compute  $\Pi_{\text{III}}$,
\begin{eqnarray}
 \Pi_{\text{III}}    =-\frac{1}{2}\left(\frac{\partial}{\partial m^2}\right)^2  T\sum_n \int \frac{d^3p}{(2\pi)^3}\frac{(qE)^2+(qB)^2}{\omega_n^2+p^2+m^2}, \nonumber \\
\end{eqnarray}
where we have employed
\begin{equation}
    \frac{(-1)^n}{n!} \left(\frac{\partial}{\partial m^2}\right)^n \frac{1}{\omega_n^2+p^2+m^2}=\frac{1}{(\omega_n^2+p^2+m^2)^{n+1}}. \label{derivada}
\end{equation}
Making use of the Matsubara frequencies according to Eq. (\ref{suma}), we obtain 
\begin{eqnarray}
 \Pi_{\text{III}}&=&-\frac{(qE)^2+(qB)^2}{2}\left(\frac{\partial}{\partial m^2}\right)^2\frac{1}{2\pi^2}\int_0^{\infty}dp \frac{p^2}{2 \sqrt{p^2+m^2}} \nonumber \\
 &&\times(1+2n_B(\sqrt{p^2+m^2})).
\end{eqnarray} 
We notice that the first term in the previous expression does not correspond to the vacuum, as it was the case in  $\Pi_{\text{I}}$. Here we have pure electric and magnetic contributions.  The integral can be done using Eq. (\ref{bessel}) getting
\begin{eqnarray}
 \Pi_{\text{III}}&=&-((qE)^2+(qB)^2)\Biggl[\frac{1}{32 \pi^2 m^2} \nonumber \\
 &+& \frac{1}{16\pi^2mT}\sum_{n=1}^{\infty}n K_1(nm/T) \Biggr].
 \end{eqnarray} 
In order to calculate $\Pi_{\text{IV}}$ we use first Eq. (\ref{derivada}) with $n=3$, finding
\begin{eqnarray}
 \Pi_{\text{IV}}    &=&-\frac{1}{6}\left(\frac{\partial}{\partial m^2}\right)^3 \nonumber \\
 &\times&T\sum_n \int \frac{d^3p}{(2\pi)^3}2\frac{(qB)^2p_{\perp}^2+(qE)^2(p_2^2+p_3^2)}{(\omega_n^2+p^2+m^2)}. \nonumber \\
\end{eqnarray}
Due to the symmetry of the integral we can replace $p_{\perp}^2=2p/3$ and $p_2^2+p_3^2=2p/3$, having then
\begin{eqnarray}
 \Pi_{\text{IV}}    &=&-\frac{2((qB)^2+(qE)^2)}{9}\left(\frac{\partial}{\partial m^2}\right)^3 \nonumber \\
 &\times&T\sum_n \int \frac{d^3p}{(2\pi)^3}\frac{p^2}{(\omega_n^2+p^2+m^2)}. \nonumber \\
\end{eqnarray}
After the sum over Matsubara frequencies we get 
\begin{eqnarray}
 \Pi_{\text{IV}}    &=&-\frac{((qB)^2+(qE)^2)}{9\pi^2}\left(\frac{\partial}{\partial m^2}\right)^3\int_0^{\infty}dp \frac{p^4}{2 \sqrt{p^2+m^2}} \nonumber \\
 &&\times(1+2n_B(\sqrt{p^2+m^2})). \nonumber \\
\end{eqnarray}
 In an analogous way to the case $ \Pi_{\text{III}}$, the purely electric and magnetic cases are obtained by calculating the integrals whereas for the temperature dependent case we use Eq. (\ref{bessel}), obtaining
\begin{eqnarray}
 \Pi_{\text{IV}} &=& ((qB)^2+(qE)^2) \Biggl[\frac{1}{48 \pi^2 m^2} \nonumber \\
 &-& \frac{1}{24\pi^2mT}\sum_{n=1}^{\infty}n K_1(nm/T) \Biggr].
\end{eqnarray}
Finally, we calculate  $\Pi_{\text{V}}$ by means of the same procedure as in the previous cases, i.e. using Eq. (\ref{derivada}) proceeding then with the sum over  Matsubara frequencies. We obtain 
\begin{eqnarray}
 \Pi_{\text{V}}    &=&\frac{(qE)^2m^2}{6\pi^2}\left(\frac{\partial}{\partial m^2}\right)^3\int_0^{\infty}dp \frac{p^2}{2 \sqrt{p^2+m^2}} \nonumber \\
 &&\times(1+2n_B(\sqrt{p^2+m^2})), \nonumber \\
\end{eqnarray}
and using Eq. (\ref{bessel}), we have
\begin{eqnarray}
 \Pi_{\text{V}} &=& (qE)^2m^2 \Biggl[\frac{1}{48 \pi^2 m^4} \nonumber \\
 &+& \frac{1}{48\pi^2m^2T^2}\sum_{n=1}^{\infty}n K_2(nm/T) \Biggr].
\end{eqnarray}
Putting together all different contributions, we finally obtain

\begin{eqnarray}
&&\Pi= -\frac{m^2}{16\pi^2} \biggl(
\ln\left(\frac{\widetilde{\mu}^2}{m^2}\right)+1 \biggr)+ \frac{mT}{2 \pi^2} \sum_{n=1}^{\infty}\frac{K_1(nm/T)}{n}\nonumber \\&&+\frac{[(qE)^2-(qB)^2]}{96\pi^2m^2}+\frac{(qE)^2}{48\pi^2T^2} \sum_{n=1}^{\infty}n K_2(nm/T) \nonumber \\
&-&\frac{5[(qE)^2+(qB)^2]}{48\pi^2m T} \sum_{n=1}^{\infty}n K_1(nm/T).\label{autoenergia}
\end{eqnarray}
Notice that we have separated the thermal part, the electromagnetic cotribution, the electro-thermal contribution and the thermo-magnetic contribution, respectively. Using our result for the self-energy correction, we will proceed now to to present graphs for Eq. (\ref{masacorregida}). As a reference value for the bare mass we will take the pion mass, i.e. $m_0=140~\text{MeV}$.  The idea is to show a graph with the evolution of the sigma mass in terms of  temperature and the electric and magnetic field intensities.   Since in the self-energy calculation our results are valid for the whole range of temperature values, we will make a graph for the sigma mass as function of temperature (without any restrictions) for different strengths of the magnetic and electric field strengths, staying nevertheless always in the weak sector in both cases. This is shown in   Fig. (\ref{masavst}).
The blue curve represents the temperature mass evolution in absence of electric and magnetic fields. The orange curve corresponds to the case where the electric field vanishes but where the magnetic field has the value of $0.9 m_{0}^2$. The green curve corresponds to the opposite case where the magnetic field vanishes and where the electric field has the value of $0.9 m_{0}^2$. Finally, the red curve represents the case where both fields have  an intensity of $0.9 m_{0}^2$. We observe that the mass grows, as we in fact expected, as function of temperature. In addition, we notice that when only the electric field is present the mass grows also with temperature. In the case where we only have a magnetic field, the mass diminishes for low temperature. When temperature becomes bigger than $T\approx 65 MeV$ the mass starts to grow. Finally, when both fields are present, the mass follows the same pattern. It diminishes for small temperature values, until $T\approx 75$, For temperature bigger than $T\approx 75 MeV$, the fields induce an a less pronounces increasing behavior for the mass. \\

\begin{figure}[h]
\includegraphics[width=8.5cm]{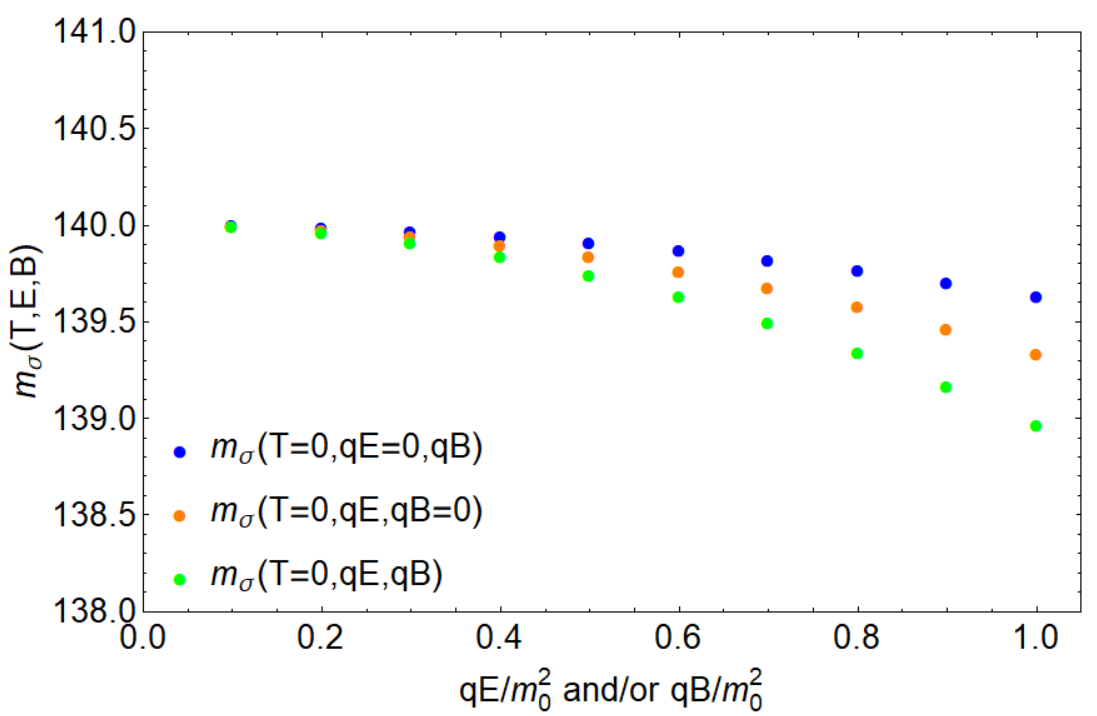} 
\caption{Color on-line. Sigma mass as function of the electric and/or magnetic field for the $T=0$ case.}
\label{masavscampo}
\end{figure}

In order to get a different perspective, we show in  Fig. (\ref{masavscampo}) a graph for the mass evolution  as function of the strengths of the fields in the $T =0$ case. The  blue curve represents the evolution of the sigma mass as function of the magnetic field when the electric field vanishes whereas the opposite case, the evolution of the mass as function of the electric field when the magnetic field vanishes, is represented by the orange curve. Finally, the green curve represents the mass evolution as function of both field intensities (growing the field strengths in the same proportion). We observe that  for the zero temperature case, in the presence of  only a magnetic field (blue curve) the mass diminishes. The opposite situation happens for a pure electric field (orange curve) where the mass grows with the field intensity. Finally, for the case when both kind of fields are present and growing in the same proportion, the mass remains constant  (green curve). 

 \section{Effective Potential}\label{sec4}

As we said in the previous section, we want also to analyze the effective potential for our  $\lambda \phi ^4$ model, including the contribution of ring diagrams, in the presence of a weak electromagnetic external fields  and temperature contributions, as well, being these effects valid for the whole range of temperature values.
 
In our model, the tree level potential is given by

\begin{equation}
  V^{\text{(tree)}}= -\frac{1}{2}\mu^2 v^2 +\frac{1}{16}\lambda^2 v^4,
\end{equation}
The effective potential at the one loop level corresponds to \cite{LeBellac,Kapusta}
\begin{eqnarray}
    V^{\text{(\text{1-loop})}}&=&\sum_{i=\sigma,\chi}\Biggl(\frac{T}{2}\sum_n \int \frac{d^3p}{(2\pi)^3} \ln[D(\omega_n,p,m_i)]^{-1}\Biggr) \nonumber\\
    &=&\sum_{i=\sigma,\chi}\Biggl( \frac{T}{2}\sum_n \int dm_i^2 \int \frac{d^3p}{(2\pi)^3} D(\omega_n,p,m_i)\Biggr) \nonumber \\ 
    &&\equiv  \frac{T}{2}\sum_n \int dm^2 \int \frac{d^3p}{(2\pi)^3} D(\omega_n,p,m),
    \label{efectivo}
\end{eqnarray}
where in the last equality we denoted $m_i$ as $m$. It is understood that the result corresponds to the sum of both contributions. Using now the propagator according to Eq. (\ref{debil}), we get

\begin{eqnarray}
 V^{\text{(\text{1-loop})}}&=&\frac{T}{2}\sum_n \int dm^2 \int \frac{d^3p}{(2\pi)^3} \Biggl[ \frac{1}{\omega_n^2+p^2+m^2}\nonumber \\
&+&\frac{4ip_2 \omega_n (qB) (qE)}{(\omega_n^2+p^2+m^2)^4} - \frac{(qE)^2+(qB)^2}{(\omega_n^2+p^2+m^2)^3} \nonumber \\ &+&  2\frac{(qB^2)p_{\perp}^2+(qE)^2(p_2^2+p_3^2)}{(\omega_n^2+p^2+m^2)^4} \nonumber \\
&+&\frac{2 (qE)^2 m ^2}{(\omega_n^2+p^2+m^2)^4}  \Biggr] \nonumber \\
&\equiv& V_{\text{I}}+V_{\text{II}}+V_{\text{III}}+V_{\text{IV}}+V_{\text{V}},
\end{eqnarray}
where
\begin{eqnarray}
    V_{\text{I}}&=&\frac{T}{2}\sum_n \int dm^2 \int \frac{d^3p}{(2\pi)^3} \frac{1}{\omega_n^2+p^2+m^2} \nonumber \\
    V_{\text{II}}&=&\frac{T}{2}\sum_n \int dm^2 \int \frac{d^3p}{(2\pi)^3}\frac{4ip_2 \omega_n (qB) (qE)}{(\omega_n^2+p^2+m^2)^4} \nonumber \\
    V_{\text{III}}&=&- \frac{T}{2}\sum_n \int dm^2 \int \frac{d^3p}{(2\pi)^3}\frac{(qE)^2+(qB)^2}{(\omega_n^2+p^2+m^2)^3} \nonumber \\
    V_{\text{IV}}&=&\frac{T}{2}\sum_n \int dm^2 \int \frac{d^3p}{(2\pi)^3}2\frac{(qB)^2p_{\perp}^2+(qE)^2(p_2^2+p_3^2)}{(\omega_n^2+p^2+m^2)^4} \nonumber \\
    V_{\text{V}}&=& \frac{T}{2}\sum_n \int dm^2 \int \frac{d^3p}{(2\pi)^3}\frac{2 (qE)^2 m ^2}{(\omega_n^2+p^2+m^2)^4}.
\end{eqnarray}
The procedure is analogous to the previous calculation. For the case  $V_{\text{I}}$ we have a vacuum contribution and an exclusively thermal part obtaining  in the $\overline{\text{MS}}$ scheme,
\begin{eqnarray}
V_{\text{I}}&=& -\frac{m^4}{64\pi^2} \biggl(
\ln\left(\frac{\widetilde{\mu}^2}{m^2}\right)+\frac{3}{2} \biggr) -\frac{m^2T^2}{2 \pi^2} \sum_{n=1}^{\infty}\frac{K_2(nm/T)}{n^2}. \nonumber \\
\end{eqnarray}
The case $V_{\text{II}}=0$ due to the integral in $dp_2$ and for the case  $V_{\text{III}}$ we have 

\begin{eqnarray}
&&V_{\text{III}}=((qE)^2+(qB)^2) \nonumber \\
&\times&\Biggl[\frac{1}{64\pi^2}\ln\left(\frac{\widetilde{\mu}^2}{m^2}\right) + \frac{1}{16\pi^2}  \sum_{n=1}^{\infty}K_0(nm/T) \Bigg].
\end{eqnarray}
Using the same strategy employed for the term  $\Pi_{\text{IV}}$, for $V_{\text{IV}}$ we obtain
\begin{eqnarray}
&&V_{\text{IV}}=-((qE)^2+(qB)^2) \nonumber \\
&\times&\Biggl[\frac{1}{96\pi^2}\ln\left(\frac{\widetilde{\mu}^2}{m^2}\right) - \frac{1}{144 \pi^2}- \frac{1}{24\pi^2}  \sum_{n=1}^{\infty} K_0(nm/T) \Bigg]. \nonumber \\
\end{eqnarray}
Finally, for  $V_{\text{V}}$, we get
\begin{eqnarray}
&&V_{\text{V}}=-(qE)^2 \Biggl[\frac{1}{96\pi^2}\ln\left(\frac{\widetilde{\mu}^2}{m^2}\right) +\frac{1}{96\pi^2} \nonumber \\
&+&\frac{1}{24\pi^2} \sum_{n=1}^{\infty}\frac{K_0(nm/T)}{n} +\frac{m}{48\pi^2 T} \sum_{n=1}^{\infty}K_1(nm/T) \Biggr].\nonumber \\
\end{eqnarray}
The sum of all contributions gives us finally the one loop effective potential

\begin{eqnarray}
&&V^{\text{(\text{1-loop})}}=\sum_{i=\sigma,\chi}\Biggl( -\frac{m_i^4}{64\pi^2} \biggl(
\ln\left(\frac{\widetilde{\mu}^2}{m_i^2}\right)+\frac{3}{2} \biggr) \nonumber \\
&&-\frac{m_i^2T^2}{2 \pi^2} \sum_{n=1}^{\infty}\frac{K_2(nm_i/T)}{n^2}\nonumber \\
&&+\frac{(qB)^2-(qE)^2}{192\pi^2}\ln\left(\frac{\widetilde{\mu}^2}{m_i^2}\right) \nonumber \\ &&+\frac{5[(qE)^2+(qB)^2]}{48\pi^2} \sum_{n=1}^{\infty}n K_0(nm_i/T) \nonumber \\
&+&(qE)^2\Biggl[-\frac{1}{288\pi^2} - \frac{1}{24\pi^2} \sum_{n=1}^{\infty}\frac{K_0(nm_i/T)}{n} \nonumber \\
&& -\frac{m_i}{48\pi^2 T} \sum_{n=1}^{\infty}K_1(nm_i/T)\Biggr] +\frac{(qB)^2}{144\pi^2} \Biggr).\label{potencial}
\end{eqnarray}
\subsection{The ring potential}
\begin{figure}[h]
\includegraphics[width=8.5cm]{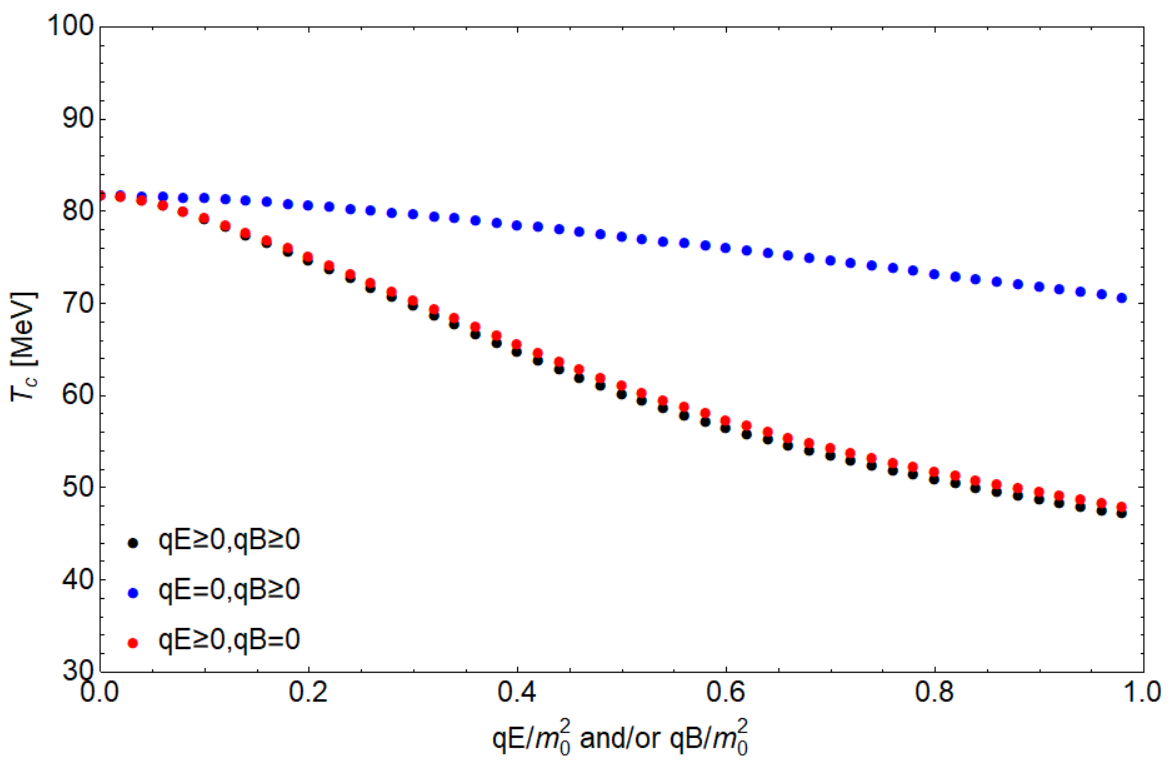} 
\caption{Color on-line. Critical temperature as function of the fields. The  black curve corresponds to the case where both fields are present. The red curve is for the case where we have only an electric field and the blue curve corresponds to the case where only a magnetic field is present.  }
\label{tcritica}
\end{figure}

The ring contribution is given by  \cite{LeBellac,Kapusta}
\begin{equation}
  V^{\text{(\text{ring})}}=  \frac{T}{2}\sum_n \int \frac{d^3p}{(2\pi)^3} \ln[1+ \Pi(\omega_n,p)D(\omega_n,p)].
\end{equation}
The previous expression can be written as 
\begin{eqnarray}
  V^{\text{(\text{ring})}}&=&  \frac{T}{2}\sum_n \int \frac{d^3p}{(2\pi)^3} \ln\bigl[(D(\omega_n,p)^{-1}+\Pi(\omega_n,p)) \nonumber \\
  &\times& (D(\omega_n,p))\bigr] \nonumber \\
  &=&\frac{T}{2}\sum_n \int \frac{d^3p}{(2\pi)^3} \ln[D(\omega_n,p)] \nonumber \\
  &+&\frac{T}{2}\sum_n \int \frac{d^3p}{(2\pi)^3} \ln[D(\omega_n,p)^{-1}+\Pi(\omega_n,p)],\nonumber \\
\end{eqnarray}
Notice that by considering the sum of the one loop and  rings contributions, we get
\begin{eqnarray}
 V^{\text{(\text{1-loop})}}+V^{\text{(\text{ring})}}&=& \frac{T}{2}\sum_n \int \frac{d^3p}{(2\pi)^3} \ln\bigl[D(\omega_n,p)^{-1} \nonumber \\
 &+&\Pi(\omega_n,p)\bigr]. 
\end{eqnarray}
Therefore, it is not necessary to calculate something else. In fact, we have just to replace the mass  $m^2$ by $m^2+\Pi$ in  Eq. (\ref{potencial}). The expression for the effective potential up to  ring order is given by

\begin{eqnarray}
&&  V^{\text{(tree)}}+V^{\text{(\text{1-loop})}}+V^{\text{(\text{ring})}}=\nonumber \\
&& -\frac{1}{2}\mu^2 v^2 +\frac{1}{16}\lambda^2 v^4 \nonumber \\
&+&\sum_{i=\sigma,\chi}\Biggl( -\frac{(m_i^2+\Pi_i)^2}{64\pi^2} \biggl(
\ln\left(\frac{\widetilde{\mu}^2}{m_i^2+\Pi_i}\right)+\frac{3}{2} \biggr) \nonumber \\
&&-\frac{(m_i^2+\Pi_i)T^2}{2 \pi^2} \sum_{n=1}^{\infty}\frac{K_2(n\sqrt{m_i^2+\Pi_i}/T)}{n^2}\nonumber \\
&&+\frac{(qB)^2-(qE)^2}{192\pi^2}\ln\left(\frac{\widetilde{\mu}^2}{m_i^2+\Pi_i}\right) \nonumber \\ &&+\frac{5[(qE)^2+(qB)^2]}{48\pi^2} \sum_{n=1}^{\infty}n K_0(n\sqrt{m_i^2+\Pi_i}/T) \nonumber \\
&+&(qE)^2\Biggl[-\frac{1}{288\pi^2} - \frac{1}{24\pi^2} \sum_{n=1}^{\infty}\frac{K_0(n\sqrt{m_i^2+\Pi_i}/T)}{n} \nonumber \\
&& -\frac{m_i}{48\pi^2 T} \sum_{n=1}^{\infty}K_1(n\sqrt{m_i^2+\Pi_i}/T)\Biggr] +\frac{(qB)^2}{144\pi^2} \Biggr),\label{potencialring}
\end{eqnarray}

where $\Pi_i$ is given by Eq. (\ref{autoen}). We stress that the inclusion of rings is crucial for the analyticity of the effective potential, avoiding the appearance of imaginary mass terms. If we want to determine the critical temperature where the symmetry is restored, we need to explore the behavior of the effective potential as function of the expectation value $(v)$. It becomes more flat for a growing temperature. At the critical temperature the first and second derivatives, in fact, all derivatives, vanish becoming then convex for $T > T_{c}$. Here we have to do with a second order phase transition since in the whole procedure no degenerated vacuums appear. We obtain, then, the behavior of the critical temperature as function of the strengths of the electromagnetic fields which is shown in  Fig. (\ref{tcritica}). The blue curve corresponds to the case where we only have an external magnetic field finding inverse magnetic catalysis (IMC), i.e. the situation where the critical temperature diminishes as function of the magnetic field intensity. The red curve corresponds to the case of a pure electric field,  having inverse electric catalysis (IEC), i.e. a scenario where the critical temperature diminishes as function of the electric field strength. The orange curve corresponds to the situation where both fields are present showing the occurrence of inverse magnetic-electric catalysis (IMEC). In this case, both fields cooperate for the occurrence of IMEC.

\section{Conclusions}\label{sec5}

The main novelty of this work is the derivation of a bosonic propagator in a proper time representation, when external constant magnetic an electric fields are simultaneously present in a perpendicular configuration.  Since this propagator is quite cumbersome, in order to  get analytical results we have presented a weak field approximation which makes possible to get analytic results for  mass corrections and for the effective potential due to the electric and magnetic field effects, including also temperature in the discussion. We worked in the frame of a complex scalar self-interacting field theory. Our results for the mass correction were presented  in  figures (\ref{masavst}) and (\ref{masavscampo}). The mass grows with temperature when we also turn the electric field on. When the magnetic field is turned on, however,the mass diminishes for small temperature values (up to $T\approx 65 MeV$). For higher temperature values, it starts again to grow. For the situation where both fields are present, the mass also diminish for small temperatures (up to  $T\approx 75 MeV$) having a less pronounced growing behavior for higher temperatures. From Fig. (\ref{masavscampo}) When $T=0$ we see that the mass diminishes as function of the magnetic field strength, growing with the intensity of the electric field. These opposite behaviors can also be observed in other scenarios as, for example, the determination of scattering lengths or the dependence of renormalon residues \cite{renormalon0,zamora5,scatering1,scatering2}.

As an important conclusion about the effective potential, we want to stress the occurrence of IMC or IEC for the cases where only a magnetic field or only an electric field are present, respectively. IMEC was clearly found, when both fields are simultaneously present, cooperating both fields for the IMEC effect. Although we have used a quite simple model, the central result, namely the explicit propagator in the presence of external electric and/or magnetic fields, could be used in more realistic scenarios as, for example, the linear sigma model. We would like to explore this case in a future work.

\section*{Acknowledgements}
M. Loewe and R. Zamora acknowledge support from   ANID/CONICYT FONDECYT Regular (Chile) under Grants No. 1200483 and 1220035. M.L.  acknowledges support from FONDECYT Regular under grant No. 1190192. ML acknowledges also support from ANID PIA/APOYO AFB 180002 (Chile) and from the Programa de de Financiamiento Basal FB210008 para Centros Cient\'ificos y Tecnológicos de excelencia de ANID.







\newpage
\begin{thebibliography}{89}

%
\bibitem{iranianos} Sh. Fayazbakhsh and N. Sadooghi, Phys. Rev. D {\bf 88}, 065030 (2013).

\bibitem{zamora1} A. Ayala, J.L Hern\'andez, L. A. Hern\'andez, R.L.S Farias and R. Zamora, Phys. Rev. D 103, 054038 (2021).

\bibitem{zamora2} A. Ayala, J.L. Hern\'andez, L. A. Hern\'andez, R.L.S Farias and R. Zamora, Phys. Rev. D 102, 114038 (2020).

\bibitem{zamora3} C.A. Dominguez, L. A. Hern\'andez, M. Loewe, C. Villavicencio and   R. Zamora, Phys. Rev. D 102, 094007 (2020).
%
\bibitem{simonov03} Yu. A. Simonov, Phys. At. Nucl. {\bf 79}, 455 (2016).
%
\bibitem{aguirre02} R.~M.~Aguirre,
Eur. Phys. J. A \textbf{55}, 28 (2019).
%
\bibitem{tetsuya} T. Yoshida and K. Suzuki,
Phys. Rev. D {\bf 94}, 074043 (2016).
%
\bibitem{dudal04} D. Dudal and T. G. Mertens,
Phys. Rev. D {\bf 91}, 086002 (2015).
%
\bibitem{kevin} K.~Marasinghe and K.~Tuchin,
Phys. Rev. C {\bf 84}, 044908 (2011).
%
\bibitem{gubler} P. Gubler, K. Hattori, S. H.ff Lee, M. Oka, S. Ozaki and K. Suzuki, Phys. Rev. D {\bf 93}, 054026 (2016).
%
\bibitem{noronha01} C. S. Machado, S. I. Finazzo, R. D. Matheus and J. Noronha, Phys. Rev. D {\bf 89}, 074027 (2014).
%
\bibitem{morita} S. Cho, K. Hattori, S. H. Lee, K. Morita and S. Ozaki,
Phys. Rev. Lett. {\bf 113}, 172301 (2014).


\bibitem{Ayala1} A. Ayala, R. L. S. Farias, S. Hern\'andez-Ort\'iz, L. A. Hern\'andez, D. Manreza Paret and R. Zamora, Phys. Rev. D {\bf 98}, 114008 (2018).
%
\bibitem{morita02}  S. Cho, K. Hattori, S. H. Lee, K. Morita and S. Ozaki,
Phys. Rev. D {\bf 91}, 045025 (2015).
%
\bibitem{sarkar03} S.~Ghosh, A.~Mukherjee, M.~Mandal, S.~Sarkar and P.~Roy,
Phys. Rev. D \textbf{94}, 094043 (2016).
%
\bibitem{band} A.~Bandyopadhyay and S.~Mallik,
Eur. Phys. J. C \textbf{77}, 771 (2017).
%

%
\bibitem{nosso1} S. S. Avancini, W. R. Tavares and M. B. Pinto, Phys. Rev. D {\bf 93}, 014010 (2016).
%
\bibitem{nosso03} S. S. Avancini, R. L. Farias, M. B. Pinto, W. R. Tavares and V. S. Timteo, Phys. Lett. B {\bf 767}, 247 (2017).



\bibitem{Ayala2} A. Ayala, C. A. Dominguez, L. A.
Hern\'andez, M. Loewe and R. Zamora, Phys. Rev. D {\bf 92}, 096011 (2015).


\bibitem{zamora4} A. Ayala, S. Hernandez-Ortiz, L. A. Hernandez, V. Knapp-Perez and R. Zamora, Phys. Rev. D 201, 074023 (2020).



\bibitem{peng}
M. Ruggieri and G. X. Peng, Phys. Rev. D 93, 094021 (2016).

\bibitem{cohen}
D. Cohen, D. A. McGady, and E. S. Werbos, Phys. Rev. C 76, 055201 (2007).

\bibitem{tavares}
W. R. Tavares and S. S. Avancini, Phys. Rev. D 97, 094001 (2018).


\bibitem{inverse3}
Pedro Costa, Márcio Ferreira, Débora P. Menezes, Jo\~ao
Moreira, and Constança Provid\^encia, Phys. Rev. D,
92, 036012 (2015).

\bibitem{inverse4}
Jens O. Andersen, Eur. Phys. J. A (2021) 57: 189.

\bibitem{inverse5}
A. Ahmad and A. Raya, J. Phys. G43, 065002, 2016.

\bibitem{inverse6}
M. Ferreira, P. Costa, O. Lourenç0, T. Fredetico and C. Provid\^encia, Phys. Rev. D,
89, 116011 (2014).

\bibitem{inverse7}
A. Ayala, L. A. Hernández, M. Loewe, J. C. Rojas and R. Zamora, Eur. Phys. J. A56, (2020).



\bibitem{reviews}. V. A. Miransky and I. A. Shovkovy, Phys. Rept. 576, 1 (2015);  D. Kharzeev, K. Landsteiner, A. Schmitt, Ho-Ung Yee, Editors, Strongly Interacting Matter in
Magnetic Fields, © Springer-Verlag Berlin Heidelberg 2013, and references therein. G. Baym.T. Hatsuda, T. Kojo, P.D. Powell, Y. Song, and T. Takasuka, Rept. Prog. Phys. 81 (2018) 5, 056902; D. Blaschke, A. Ayriyan, and A. Friesen (Editors), ``Compact Stars in the QCD Phase Diagram", Universe, MppiaG (2020). 

\bibitem{lattice}
G. S. Bali, F. Bruckmann, G. Endr{\"o}di, Z. Fodor, S. D. Katz and A. Sch{\"a}fer, Phys. Rev. D {\bf 86}, 071502 (2012);  G. Bali, F. Bruckmann, G. Endr{\"o}di, Z. Fodor, S. Katz, S. Krieg, A. Schaefer, and K. K. Szabo, JHEP
1202, 044 (2012);  G. Bali, F. Bruckmann, G. Endrödi, S. Katz, and A. Sch{\"a}fer, JHEP 1408, 177 (2014).  

\bibitem{nosotros1}
A. Ayala, M. Loewe and R. Zamora, Phys. Rev. D {\bf91}, 016002 (2015); 

\bibitem{nosotros2}
A. Ayala, M. Loewe, A. Mizher and R. Zamora, Phys. Rev. D {\bf90}, 036001 (2014).


\bibitem{cristian} Alejandro Ayala, Luis. A. Hern\'andez, Marcelo Loewe, and Cristian Villavicencio, Eur. Phys. J.A 57 (2021), 7, 234. 

\bibitem{ultimonosotros}
M. Loewe, D. Valenzuela and R. Zamora, Phys. Rev. D {\bf 105}, 036017 (2022).

\bibitem{Farias} W. R. Tavares, R. L. S. Farias and S. S. Avancini, Phys. Rev. D {\bf 101}, 016017 (2020).

\bibitem{renormalon2}
M. Loewe and R. Zamora, Phys. Rev. D {\bf 105},076011 (2022).

\bibitem{paralelos}
G. Cao and X.G. Huang, Phys. Rev.
D {\bf 93}, 016007 (2016); M. Ruggeri, and G. X. Peng, Phys. Rev. D {\bf 93}, 094021 (2016); M. Ruggeri, Z. Y. Lu and G. X. Peng, Phys. Rev. D {\bf 94}, 116002 (2016).

\bibitem{Dittrich} Walter Dittrich and Martin Reuter, ``Efective Lagrangiand in Quantum Electrodynamics". Springer-Verlag, Berlin Heidelberg New York, Tokyo (1985). See also Walter Dittrich and Holger Gies, ``Probing the Quantum Vacuum: Perturbative effective action approach in Queantum Electrodynamics and its application", Springer Tracts in Modern Physics, Volume 166 (2000)

\bibitem{ahmad}
A. Ahmad, N. Ahmadiniaz, O. Corradini, S. P. Kim and C. Schubert, Nuclear Physics B, 919 (2017).

\bibitem{ayalaeuro}
A. Ayala, J. Casta\~no, L. A. Hernandez, J. Salinas and R. Zamora, Eur. Phys. J. A57, (2021).

\bibitem{LeBellac}
M. Le Bellac, {\it Thermal Field Theory}, Cambridge University Press,
1996.

\bibitem{Kapusta}
J. I. Kapusta and C. Gale, {\it Finite-Temperature Field Theory Principles and Applications}, Cambridge University Press,
2006.

\bibitem{abramowitz}
Abramowitz, M. and Stegun, I. A. (Eds.). Handbook of Mathematical Functions with Formulas, Graphs, and Mathematical Tables, 9th printing. New York: Dover, pp. 576-579, 1972.


















\bibitem{renormalon0}
M. Correa, M. Loewe and R. Zamora, Phys. Rev. D {\bf 99}, 096024 (2019).

\bibitem{zamora5} M. Loewe, L. Monje and R. Zamora, Phys. Rev. D {\bf 104}, 016020 (2021).

\bibitem{scatering2}
M. Loewe, E. Mu\~{n}oz and R. Zamora, Phys. Rev. D {\bf 100}, 116006 (2019).

\bibitem{scatering1}
M. Loewe, L. Monje, E. Mu\~{n}oz, A. Raya and R. Zamora, Phys. Rev. D {\bf 99}, 056002 (2019).




\end{thebibliography}
\end{document}